\documentclass{pnastwo_altered}

\usepackage{amssymb,amsfonts,amsmath}
\usepackage{natbib}
\usepackage[pdftex]{graphicx}
\usepackage{url}
\bibpunct{(}{)}{,}{n}{,}{,}
\begin{document}
\title{Age-aggregation bias in mortality trends\footnote{We are submitting an abridged version of this note as a letter to the Proceedings of the National Academy of Sciences.}}
\author{
Andrew Gelman\affil{1}{Department of Statistics, Columbia University}\affil{2}{Department of Political Science, Columbia University}
\and
Jonathan Auerbach\affil{1}{Department of Statistics, Columbia University}
}
\date{25 Nov 2015}
\maketitle

In a recent article in PNAS, Case and Deaton \cite{CaseDeaton2015} show a figure illustrating ``a marked increase in the all-cause mortality of middle-aged white non-Hispanic men and women in the United States between 1999 and 2013.'' The authors state that their numbers ``are not age-adjusted within the 10-y 45-54 age group.''  They calculated the mortality rate each year by dividing the total number of deaths for the age group by the population of the age group.

We suspected an aggregation bias and examined whether much of the increase in aggregate mortality rates for this age group could be due to the changing composition of the 45--54 year old age group over the 1990 to 2013 time period. If this were the case, the change in the group mortality rate over time may not reflect a change in age-specific mortality rates. Adjusting for age confirmed this suspicion. Contrary to Case and Deaton's figure, we find there is no longer a steady increase in mortality rates for this age group. Instead there is an increasing trend from 1999--2005 and a constant trend thereafter. Moreover, stratifying age-adjusted mortality rates by sex shows a marked increase only for women and not men, contrary to the article's headline.

\begin{figure}
\centerline{\includegraphics[width=.32\textwidth]{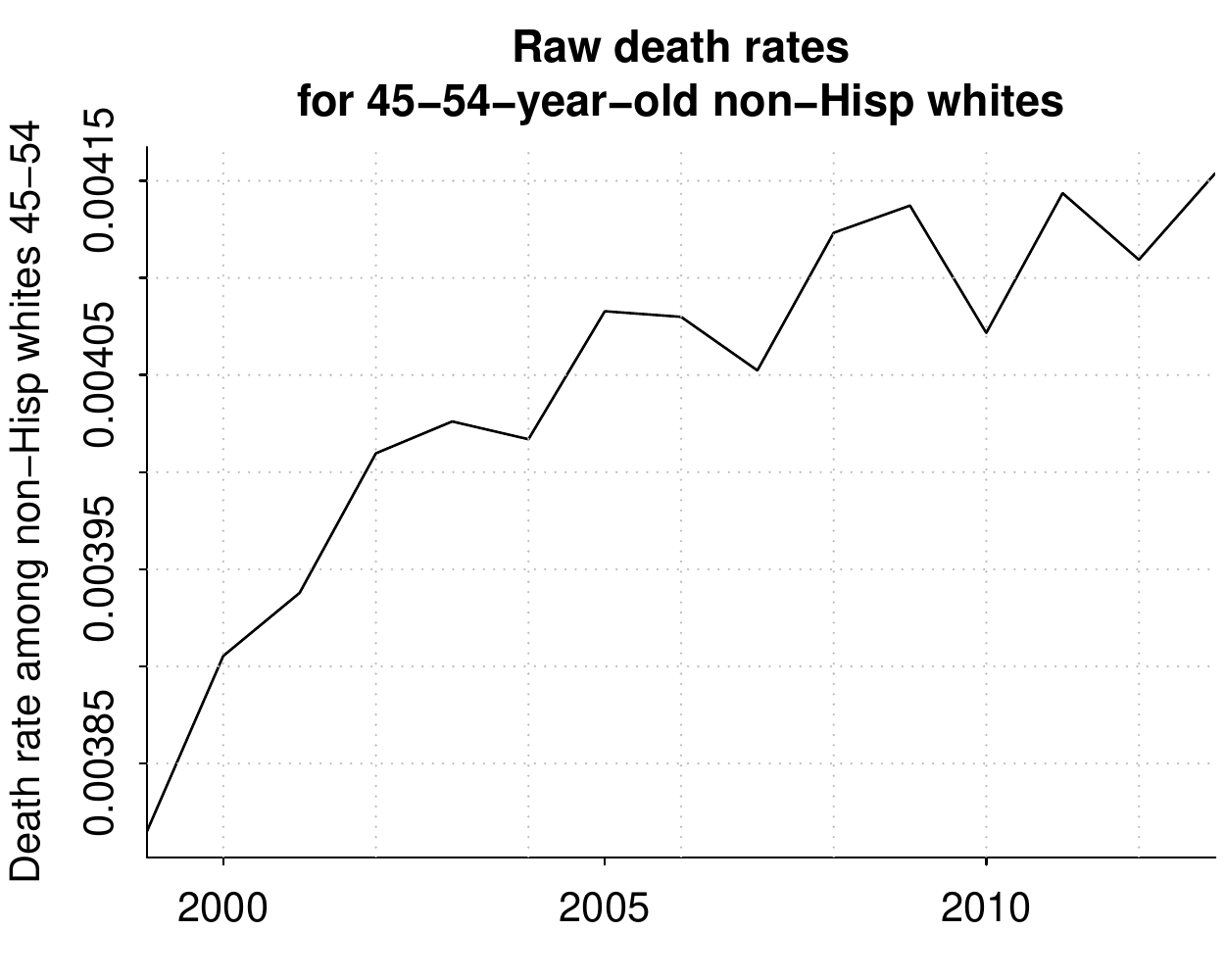}\hspace{.02\textwidth}\includegraphics[width=.32\textwidth]{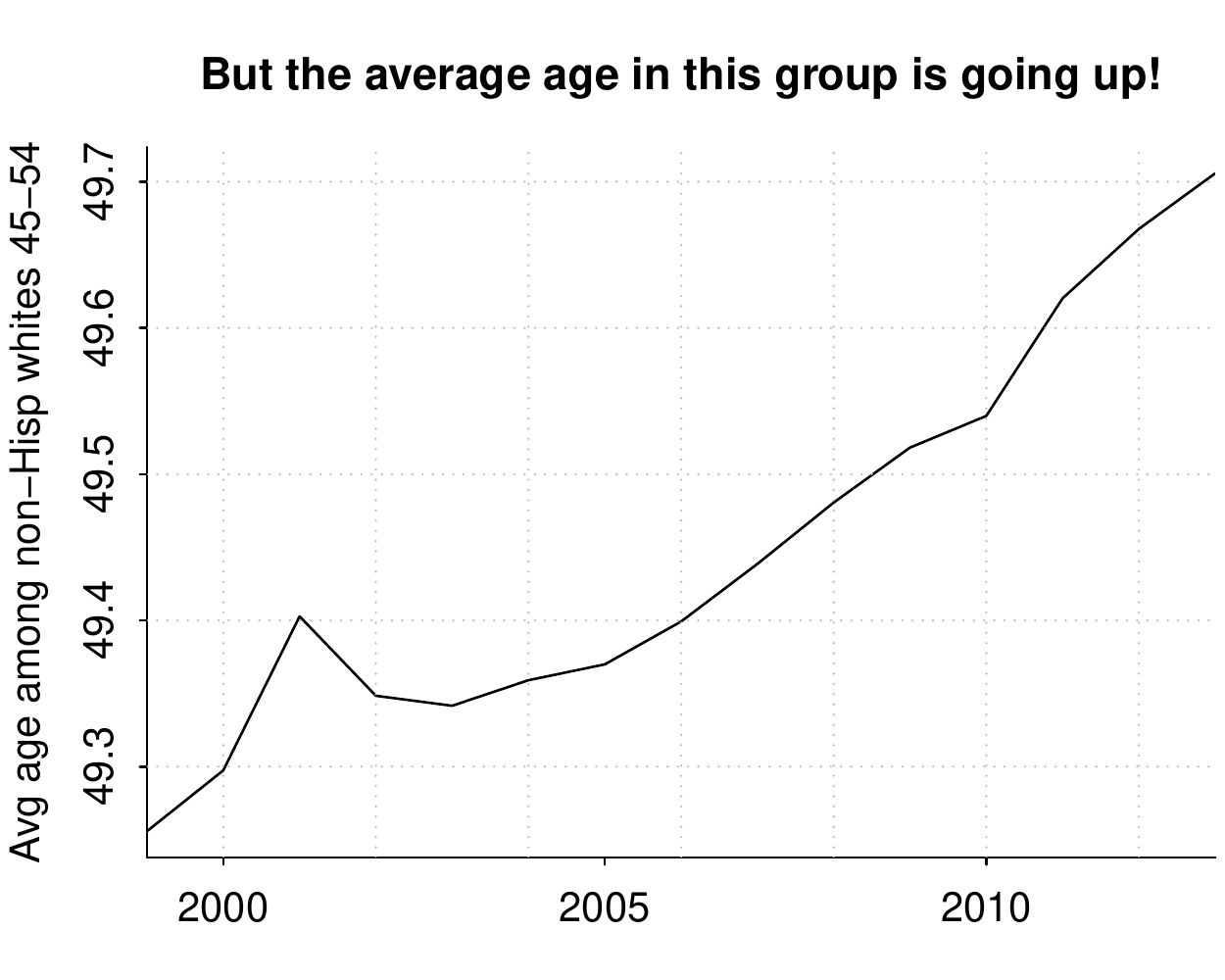}\hspace{.02\textwidth}\includegraphics[width=.32\textwidth]{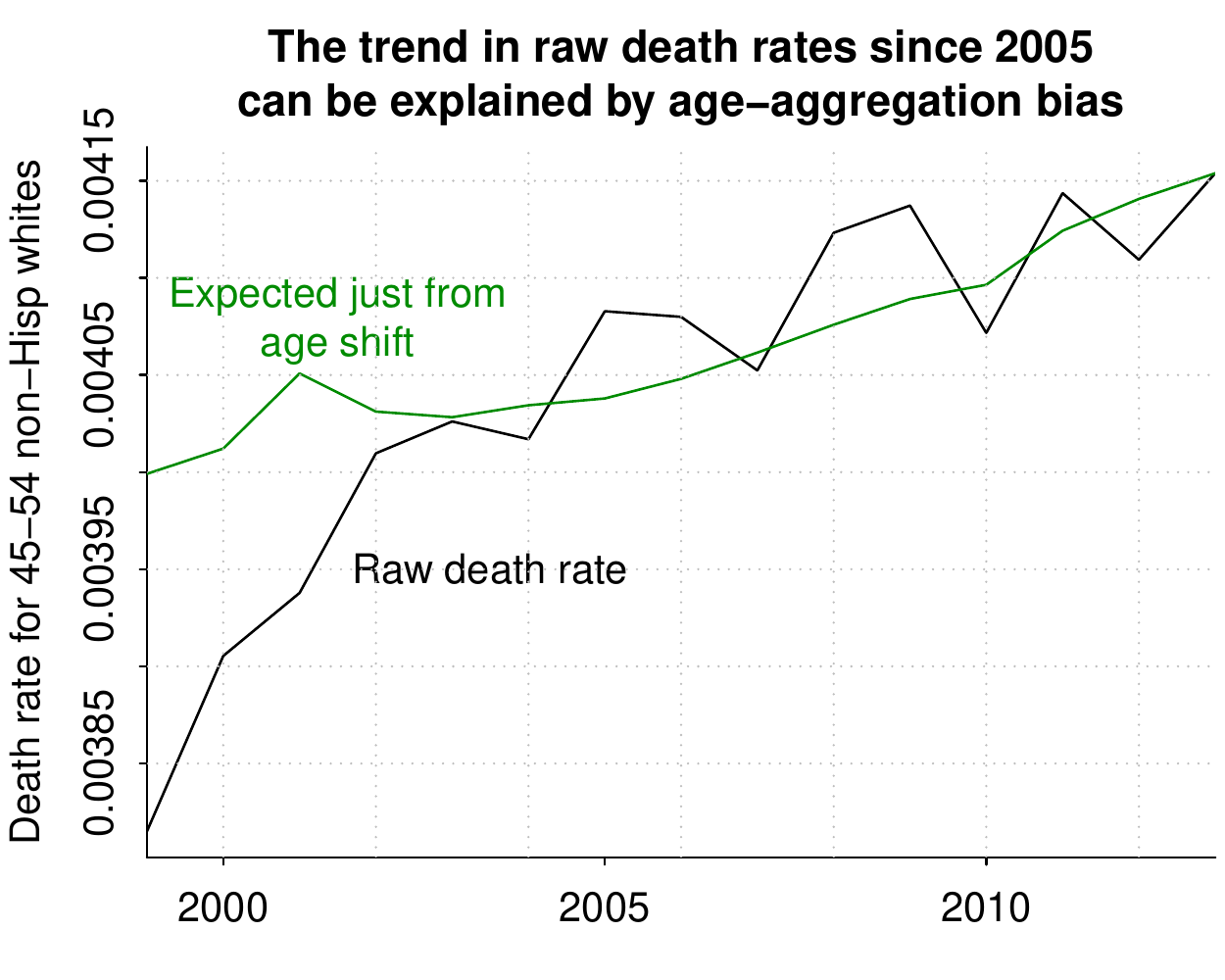}}
\caption{(a) Observed increase in raw mortality rate among 45-54-year-old non-Hispanic whites, unadjusted for age; (b) Increase in average age of this group as the baby boom generation moves through; (c) Raw death rate, along with trend in death rate attributable by change in age distribution alone, had age-specific mortality rates been at the 2013 level throughout.}\label{fig1}
\end{figure}

We demonstrate the necessity of the age adjustment in Figure \ref{fig1}. The unadjusted numbers in Figure \ref{fig1}a show a steady increase in the mortality rate of 45-54-year-old non-Hispanic whites. During this period, however, the average age in this group increased as the baby boom generation passed through. Figure \ref{fig1}b shows this increase.

Suppose for the moment that mortality rates did not change for individuals in this age group over the 1999 to 2013 time period. In this case, we could calculate the change in the group mortality rate due solely to the change in the underlying age of the population. We do this by taking the 2013 mortality rates for each age and computing a weighted average rate each year using the number of individuals in each age group. Figure \ref{fig1}c shows the result. The changing composition in age explains about half the change in the mortality rate of this group since 1999 and all the change since 2005.

Having demonstrated the importance of age-adjustment, we now adjust the numbers published in the Case and Deaton paper. We ask what the data would look like if the age groups remained the same each year and only the individual mortality rates changed. Figure 2a shows the simplest such adjustment, normalizing each year to a hypothetical uniformly-distributed population in which the number of people is equal at each age from 45 through 54. That is, we calculate the mortality rate each year by dividing the number of deaths for each age between 45 and 54 by the population of that age and then taking the average. This allows us to compare mortality rates across years. Consistent with Figure 1c, the resulting mortality rate increased from 1999 to 2005 and then stopped increasing.

We could just as easily use another age distribution to make valid comparisons across years, and we find that age-adjusted trend is not sensitive to the age distribution used to normalize the mortality rates. Figure 2b shows the estimated changes in mortality rate under three options: first assuming a uniform distribution of ages 45--54; second using the distribution of ages that existed in 1999, which is skewed toward the younger end of the 45--54 group; and third using the 2013 age distribution, which is skewed older. The general pattern does not change.

Calculating the age-adjusted rates separately for each sex reveals a crucial result, which we display in Figure 2c. The mortality rate among white non-Hispanic American women increased from 1999--2013. Among the corresponding group of men, however, the mortality rate increase from 1999--2005 is nearly reversed during 2005--2013.

In summary, age adjustment is not merely an academic exercise. Due to the changing composition of the 45-54-year-old age group, adjusting for age changes the interpretation of the data in important ways. We stress that this does not change a key finding of the Case and Deaton paper: the comparison of non-Hispanic U.S. middle-aged whites to other countries and other ethnic groups. These comparisons hold up after our age adjustment. (The aggregation bias in the published unadjusted numbers is on the order of 5 percent in the trend from 1999–-2003 while the observed changes in mortality rates in other countries and other groups are on the order of 20 percent during that period.) We suspect the comparisons would continue to hold if the remaining countries were similarly age adjusted. However, if Case and Deaton were to age-adjust these countries and stratify by gender, a more nuanced story is likely to appear.

While we do not believe that age-adjustment invalidates comparisons between countries, it does affect claims concerning the absolute increase in mortality among U.S. middle-aged white non-Hispanics. Therefore, we believe it is vital that future researchers understand the aggregation bias as they read Case and Deaton's article and consider how to investigate these noteworthy findings further.  For example, Figure \ref{fig3} shows a further breakdown of age-adjusted death rates in that group by U.S. region.  The most notable pattern has been an increase in death rates among women in the south.  In contrast, death rates for both sexes have been declining in the northeast, the region where mortality rates were lowest to begin with.  These graphs demonstrate the value of this sort of data exploration, and we are grateful to Case and Deaton for focusing attention on these mortality trends.

\bibliography{age_adj}
\pagebreak

\begin{figure}
\centerline{\includegraphics[width=.32\textwidth]{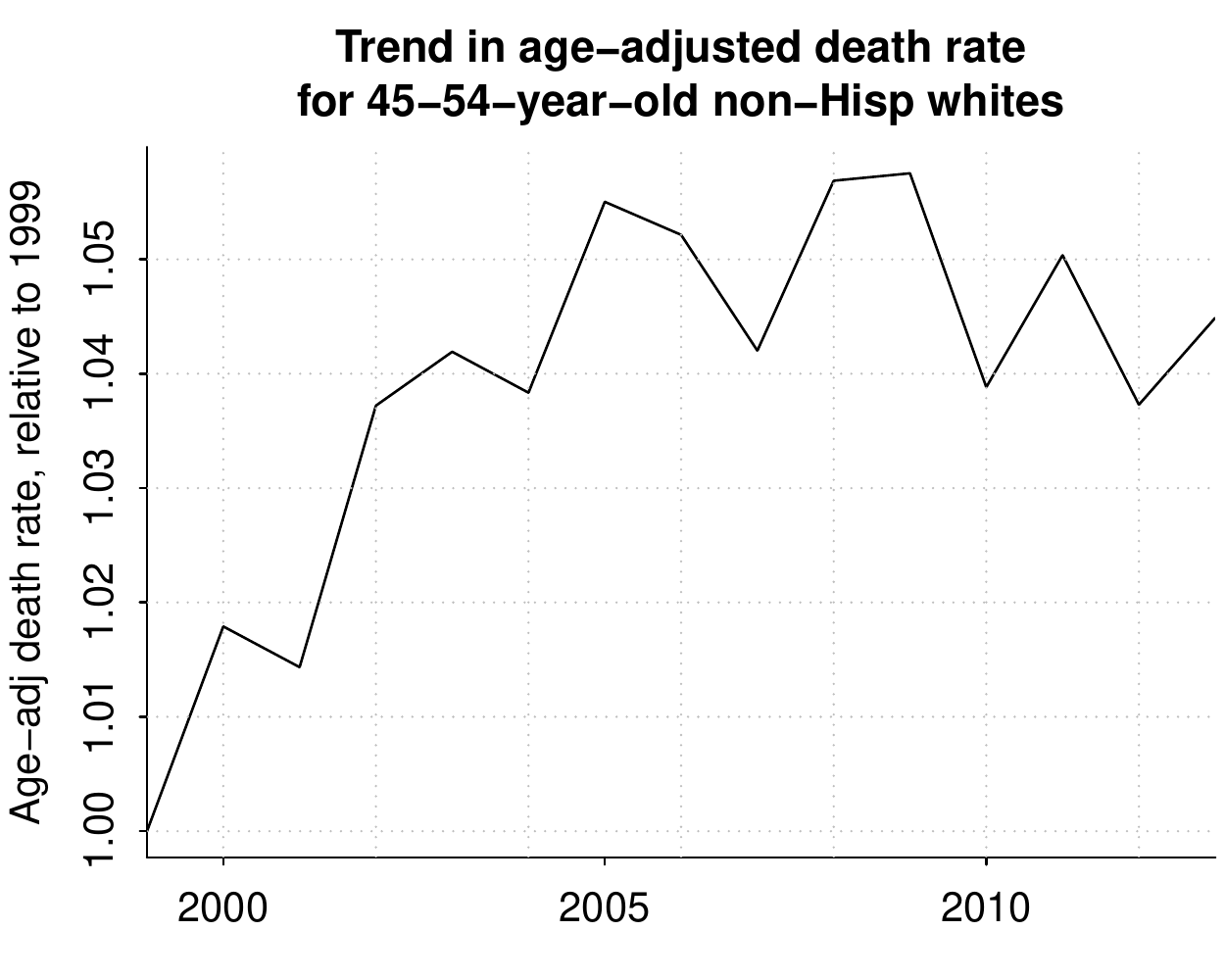}\hspace{.02\textwidth}\includegraphics[width=.32\textwidth]{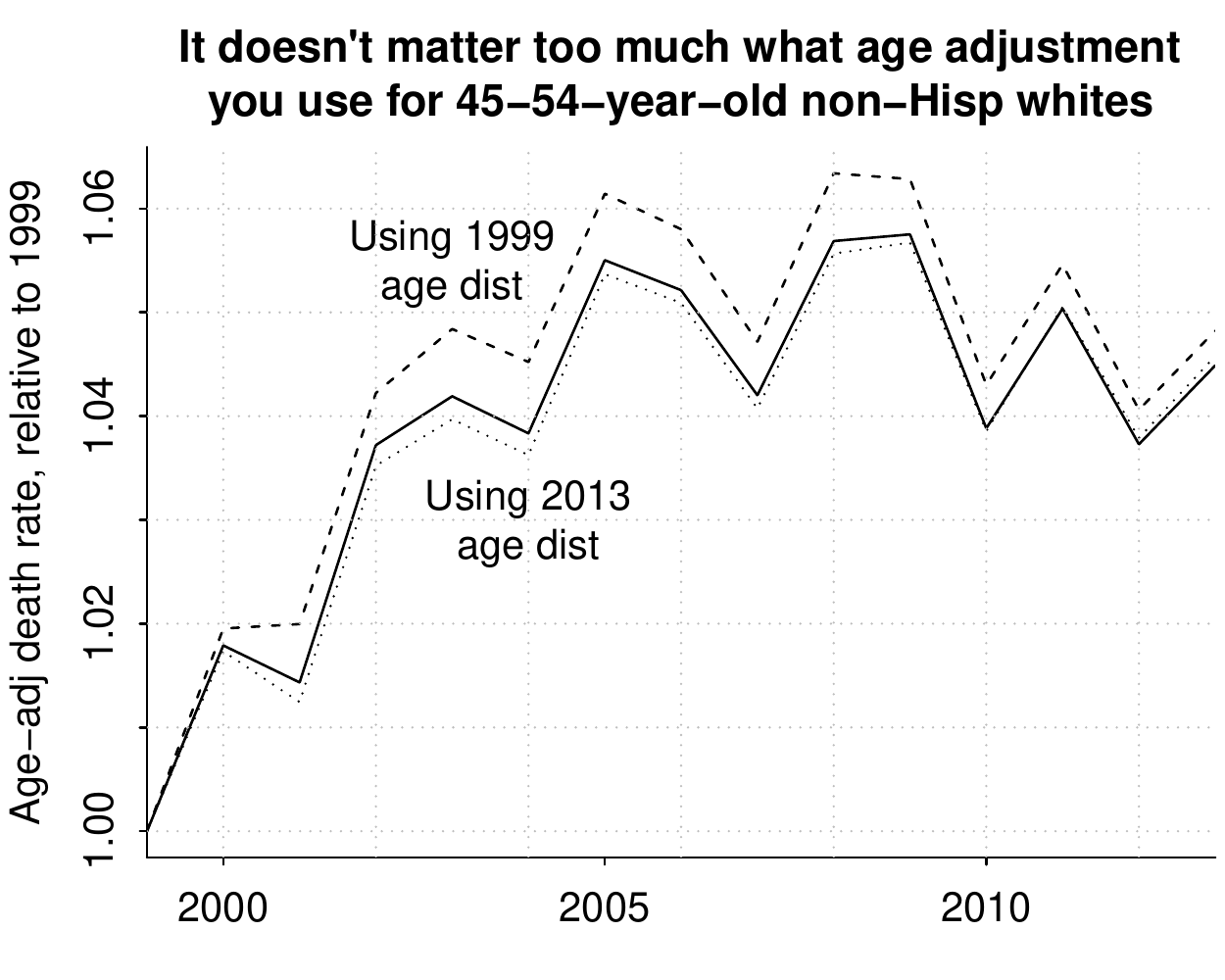}\hspace{.02\textwidth}\includegraphics[width=.32\textwidth]{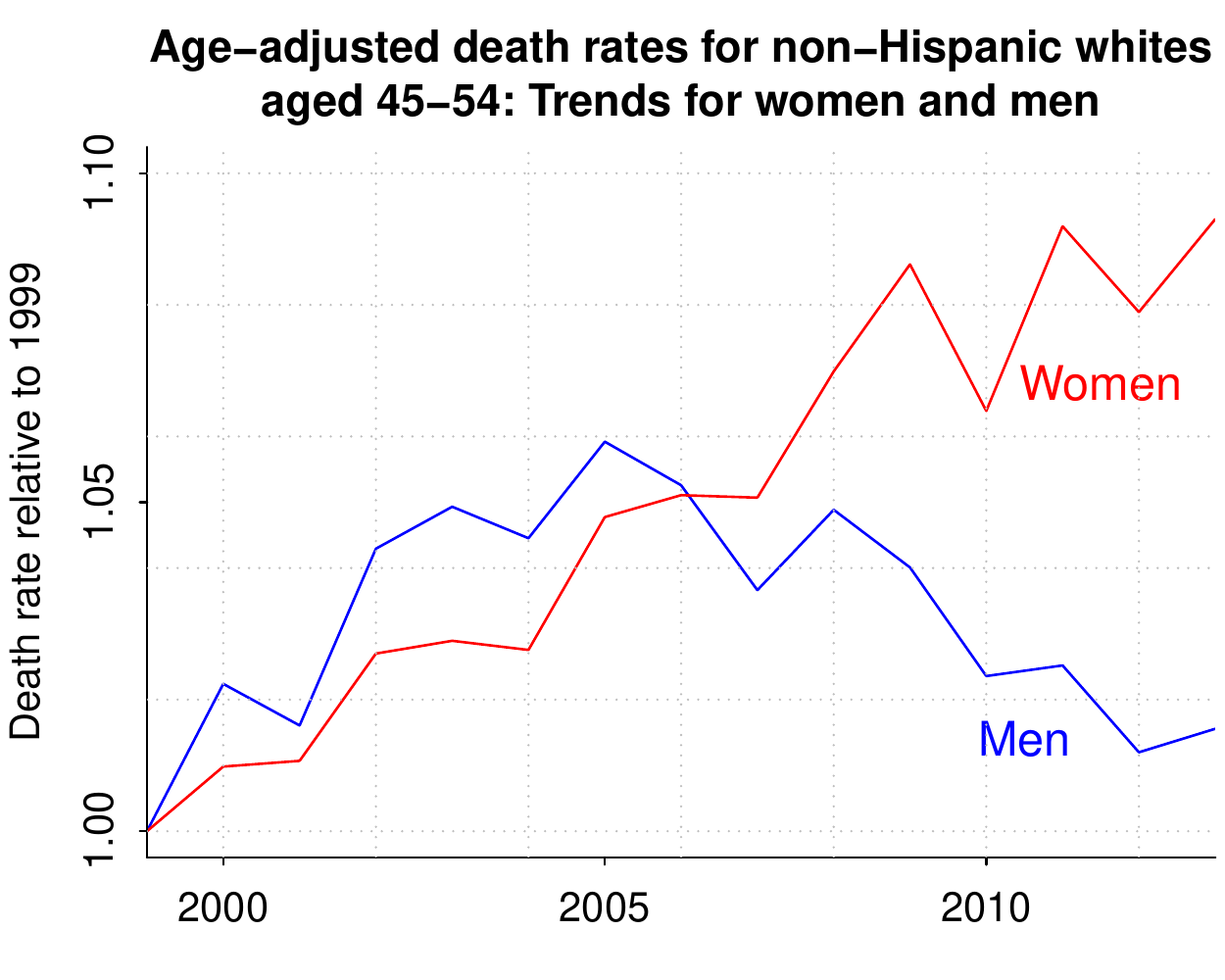}}
\caption{(a) Age adjusted death rates among 45-54-year-old non-Hispanic whites, showing an increase from 1999--2005 and a steady pattern since 2005; (b) Comparison of three different age adjustments; (c) Trends in age-adjusted death rates broken down by sex.}\label{fig2}
\end{figure}

\begin{figure}
\centerline{\includegraphics[width=.32\textwidth]{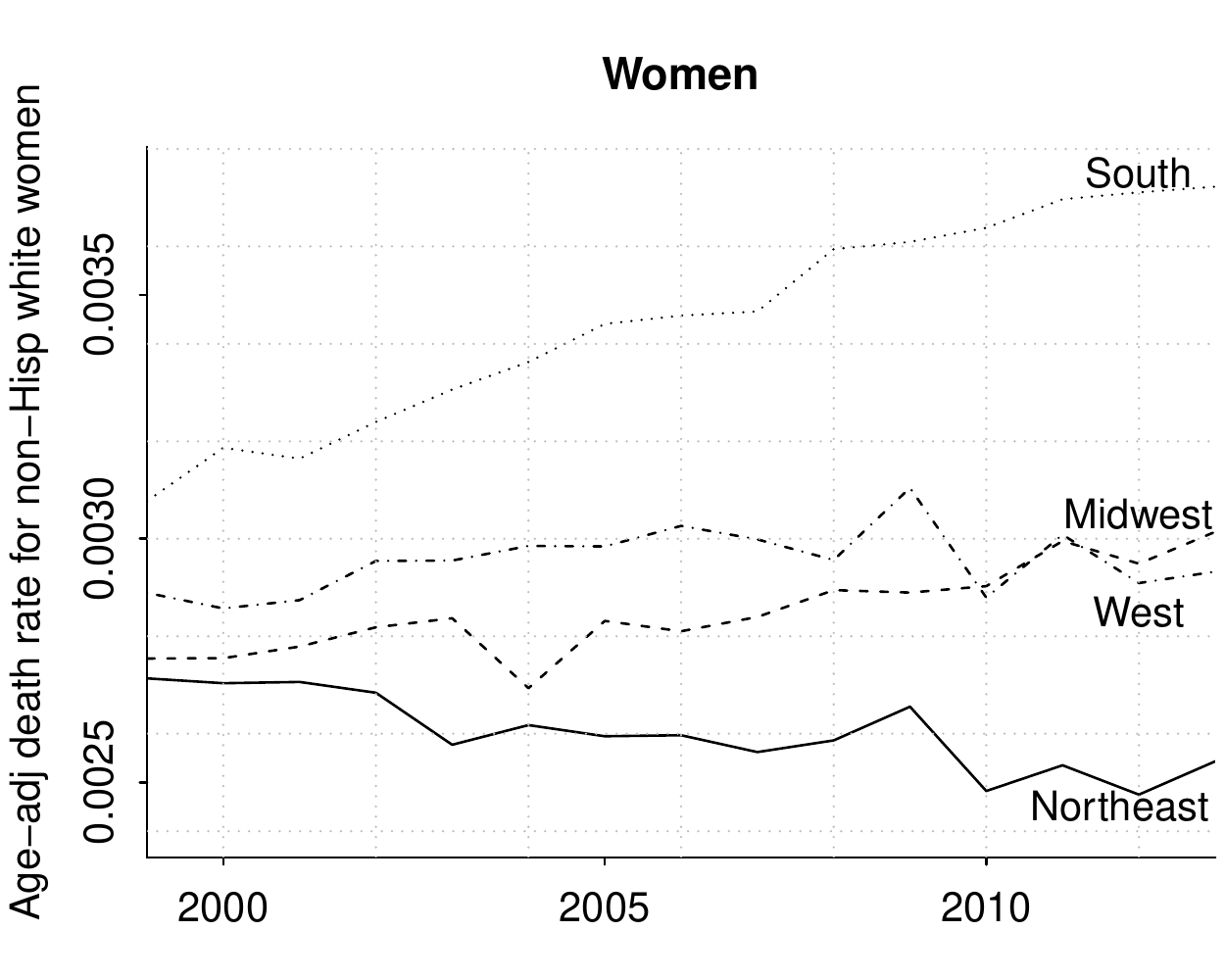}\hspace{.05\textwidth}\includegraphics[width=.32\textwidth]{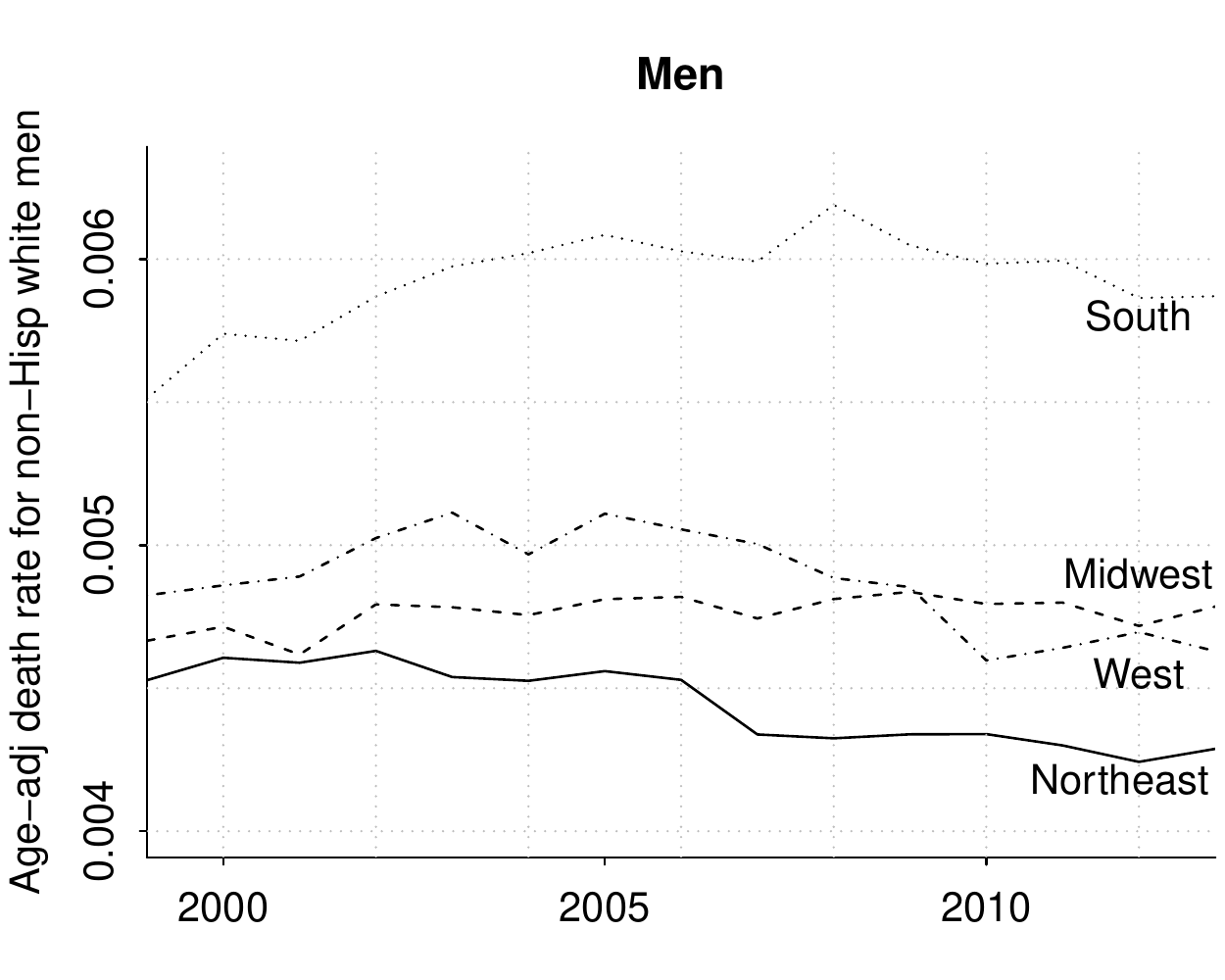}}
\caption{Age adjusted death rates among 45-54-year-old non-Hispanic white men and women, broken down by region of the country.  The most notable pattern has been an increase in death rates among women in the south.  In contrast, death rates for both sexes have been declining in the northeast.}\label{fig3}
\end{figure}

\end{document}